\documentclass{aa}

\usepackage[varg]{txfonts}

\usepackage{graphicx}
\usepackage{natbib}
\usepackage{tikz}
\usepackage{here}
\usepackage{bm}
\usepackage{url}
\usepackage{verbatim}
\usepackage{hyperref}

\hypersetup{
    colorlinks=true,
    linkcolor=blue,
    filecolor=blue,      
    urlcolor=blue,
    pdfpagemode=FullScreen,
    citecolor = blue,
    }

\begin{document}

   \title{Quasar Pairs as Large--Scale Structure tracers}


   \author{Mart\'in G. Richarte
          \inst{1}, \inst{2}\fnmsep\thanks{(martin@df.uba.ar: corresponding author)}
          Facundo Toscano\inst{3},  Diego Garcia Lambas\inst{3,4,5}, Heliana E. Luparello\inst{3}, Luiz Filipe Guimar\~aes\inst{1,6}  \and  J\'ulio C. Fabris\inst{1}}

   \institute{Departamento de F\'isica, Facultad de Ciencias Exactas y Naturales,
Universidad de Buenos Aires, Ciudad Universitaria 1428, Pabell\'on I, Buenos Aires, Argentina
\and             
N\'ucleo Cosmo-ufes \& Departamento de F\'isica - Universidade Federal do Esp\'irito Santo, 29075-910 Vit\'oria, ES, Brazil
\and 
Instituto de Astronom\'ia Te\'orica y Experimental (IATE), CONICET-UNC, C\'ordoba, Argentina
\and
Observatorio Astronómico de C\'ordoba, UNC, C\'ordoba, Argentina
\and
Comisi\'on Nacional de Actividades Espaciales (CONAE), C\'ordoba, Argentina
\and
Departamento de F\'isica, Universidade Estadual de Londrina, Rod. Celso Garcia Cid, Km 380, 86057-970 Londrina, PR, Brazil
             }


 
  \abstract
  
  {Quasars can be used as suitable tracers of the large--scale distribution of galaxies at high redshift given their high luminosity and dedicated surveys. In previous works it has been found that quasars have a bias similar to that of rich groups. Following this argument, quasar pairs could be associated with higher density environments serving as protocluster proxies.}
  {In this work, our aim is to characterize close quasar pairs residing in the same halo. This is accomplished by identifying quasar pairs in redshift space. We analyze pair--quasar cross correlations as well as quasar and quasar pairs CMB derived lensing convergence profiles centered in those systems.}
   {We identify quasar pairs in the SDSS-DR16 catalog as objects with relative velocities within $|\Delta V|\leq 900 \text{km}/s$ and projected separation less than $d_{p}\leq 2\text{Mpc}$, in the redshift range $1.2\leq z \leq 2.8$.
   We computed redshift space cross--correlation functions using Landy--Szalay estimators for the samples.
    For the analysis of the correlation between quasars/quasar pairs and the underlying mass distribution we calculate mean radial profiles of the lensing convergence parameter using Cosmic Microwave Background data provided by the Planck Collaboration.
   }
   {
   We have identified 2777 pairs of quasars in the redshift range 1.2 to 2.8. Quasar pairs show a distribution of relative luminosities that differs from that corresponding to two pairs selected at random with the same redshift distribution showing that quasars in these systems distinguish from isolated ones.
The cross--correlation between pairs and quasars show a larger correlation amplitude than the auto correlation function of quasars indicating that these systems are more strongly biased with respect to the large scale mass distribution, and reside in more massive halos. 
This is reinforced by the higher convergence CMB lensing profiles of the pairs as compared to the isolated quasars with a similar redshift distribution.
Our results show that quasar pairs are suitable precursors of present day clusters of galaxies, in contrast to isolated quasars which are associated to more moderate density environments.}

\authorrunning{Richarte, M. G. et al.} 
\titlerunning{Quasar Pairs as Large--Scale Structure tracers.}


   \maketitle
\section{Introduction}
Studies of the formation and evolution of structures in the Universe have made important recent advances from large galaxy surveys. 
Among the several new data sets, the Sloan Digital Sky Survey (SDSS) has made major contributions through the identification and redshift measurements of galaxies and quasars building homogeneous samples suitable for large-scale studies \cite{Lyke}.\\
The high luminosity of quasars make them ideal tracers of structure in the distant Universe and there have been works analysing the autocorrelation function in redshift space which allows for an estimate of the bias of these systems with respect to the mass distribution \citep[and references therein]{porci, White}. The results show the correlation function amplitude consistent with rich groups of galaxies, which goes in line with the expectation from numerical simulation models. Besides, these studies have shown a lack of luminosity dependence of the correlation function on quasar luminosity, a fact that is opposite to that found for galaxies where a strong clustering amplitude dependence on galaxy luminosity is observed \citep{Zehavi:2011}.
This independence of clustering strength on quasar luminosity is not surprising since quasar luminosity is associated to accretion onto the central black hole, a relatively stochastic process which would be relatively independent of the global environment of the quasar host when compared to the case of galaxies where luminosity produced by stars results a very good proxy of galaxy mass according to simulations \cite{noble}.\\
In this context, the exploration of the distribution of quasar systems and their association to mass may give new light on the joint formation and evolution of quasars and structure.
The relatively low number density of quasars allows to consider mainly quasar pairs as suitable systems for statistical studies, with a strongly decaying number of systems with higher membresy. \\
In this work we identify quasar pairs from the SDSS DR16 \citep{Lyke} with suitable relative separation and radial velocity. In our study we adopt a maximum projected distance of $d_{p}\leq 2\text{Mpc}$ and relative velocities within $|\Delta V|\leq 900 \text{km}/s$ as reasonable limits for the identification of physical pairs, with high probability of residing in the same halo. Our final sample of 
2777 quasar pairs is suitable for statistical studies in both correlation function and Cosmic Microwave Background (CMB) convergence lensing maps.
The selection criteria for the quasar samples are detailed in Section \ref{qso_samples}. The statistical methodologies employed to compute the autocorrelation of the quasar sample are outlined in Section \ref{2pcf}. The quasar catalog, specifically for the quasar pairs, is provided in Section \ref{QPC}. Following this, Section \ref{Lum} focuses on the characterization of luminosity and color-color maps. Section \ref{cross} presents the cross-correlation functions, while the results pertaining to the  studies on CMB convergence lensing are discussed in Section \ref{lensing}.

\section{Quasar Catalog and Subsample} \label{qso_samples}
\begin{figure}[htbp]
\includegraphics[width=\linewidth]{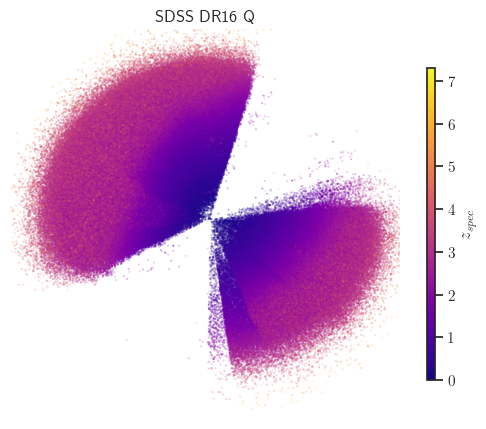}
\includegraphics[width=\linewidth]{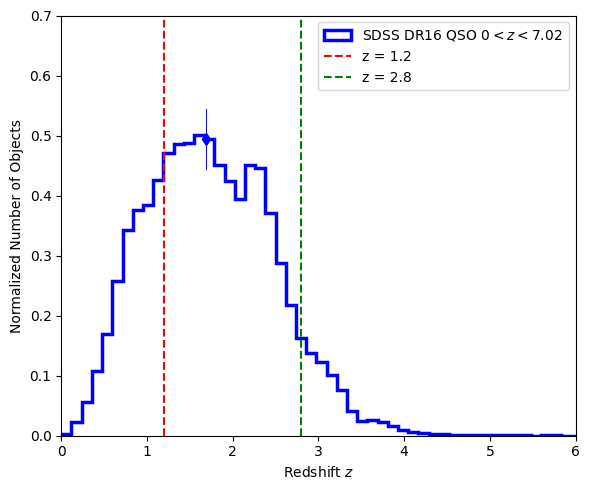}
\caption{Top panel: A projection of the 3D map  representation of the quasars from the SDSS DR16Q catalog, where the color bar indicates the redshifts of each quasar.  Bottom panel: We display the redshift distribution of the quasars, normalized to the total number of objects in the sample. The median value of this distribution is marked with a distinct point for clarity. The subsample distribution within the redshift range $z \in [1.2, 2.8]$ is highlighted with vertical lines.
\label{fig:0}} 
\end{figure}

The Sloan Digital Sky Survey (SDSS) represents a stunning multi-spectral imaging and spectroscopic redshift survey, utilizing a 2.5-meter wide-angle optical telescope in conjunction with two distinct multi-fiber spectrographs, all situated in New Mexico. One of the most recent catalog, known as SDSS DR16, has successfully confirmed the detection of approximately 750414 quasars, achieved with a spectral resolution of $R=\frac{\lambda}{\Delta \lambda} \simeq 2000$ \cite{Lyke}. The catalog spans a significant portion of cosmic history, $0<z<7.5$, and covers approximately 14555 $\text{deg}^{2}$ of the sky [see Fig. (\ref{fig:0})] . Regarding the absolute magnitude in the I band, it spans a large interval, $-30<M_{I}<-23$. The SDSS DR16 encompasses three distinct target fields: the Southern Stripe, the Northern Galactic Cap, and the Equatorial Stripe. Each of these areas has been meticulously selected to maximize the coverage of the survey and provide a comprehensive understanding of the quasar population across the sky \cite{Lyke}.\\

To enhance the detection of quasar pairs within the SDSS DR16 catalog, we constructed a sample that focuses on a more restricted redshift window of $1.2\leq z \leq 2.8$. The distribution of quasars as a function of the redshift is illustrated in Fig. (\ref{fig:0}). The analysis covers a total of 147325 quasars, which corresponds to approximately $66\%$ of the entire SDSS DR16 catalog \footnote{\url{https://www.sdss4.org/dr16/algorithms/qso_catalog}}. There are numerous compelling reasons to take into account the redshift window previously mentioned. We will delineate a series of justifications grounded in theoretical frameworks as well as numerical analyses derived from various catalogs utilized in the existing literature.  It corresponds to a pivotal epoch in the universe's history, approximately 10 to 12 billion years ago, a critical period for unraveling the processes underpinning galaxy formation and the evolution of supermassive black holes (SMBHs). During this period, the Universe underwent significant growth and activity, making it an ideal time to explore the early stages of quasar formation. Long ago, it was proposed that supermassive black holes (SMBHs) originate from the collapse of massive gas clouds in the early Universe. The energy released during the accretion process plays a crucial role in galaxy evolution, highlighting the importance of feedback mechanisms in regulating star formation within galaxies \cite{SR}.\\

Subsequently, the connection between black hole mass and the velocity dispersion of the host galaxy bulge-known as the $M_{BH}-\sigma$ relation-has been essential in showing that black hole growth is closely tied to the evolution of its host bulge. This relationship has been studied extensively in the context of hierarchical galaxy formation models, further illustrating the interconnectedness of these cosmic processes \cite{HK}, \cite{MH}. The coevolution of SMBHs and their host galaxies often results from significant mergers between galaxies, which produce feedback that affects both the black hole and the galaxy itself \cite{KH}, \cite{H}.\\

Utilizing the gravitational lensing effect, where light from a distant quasar is bent by the gravitational field of a foreground galaxy, researchers have greatly enhanced our understanding of the properties of both quasars and their host galaxies across different epochs, particularly within the redshift range $1\leq z\leq 4.5$. Analysis of 31 gravitationally lensed active galactic nuclei (AGNs) and 20 non-lensed AGNs has demonstrated the stability of the relationship between black hole mass and luminosity in the $R$-band, represented by the$M_{BH}-L_{R}$ relation, over cosmic time for $z\leq 1.7$ \cite{P}.\\
 
Additionally, a study of 11 X-ray-selected broad-line AGNs with redshifts from $1\leq z\leq 2$ suggested that black holes in the mass range of $10^{8-9}M_{\odot}$ likely formed before their host galaxies were established \cite{B}. These findings collectively highlight the complex relationship between SMBHs and their host galaxies, reinforcing the idea that the co-evolution of these systems is deeply interconnected throughout cosmic history \cite{KH}, \cite{H}.
Analysis of galaxy formation models that include AGN feedback has shown that quasars are often found in low-mass dark matter halos, typically around a few times$10^{12}M_{\odot}$ \cite{fani}. This finding suggests that these environments represent the average conditions in the low-redshift universe  ($z\leq 3$).\\

Another important aspect of quasars is how their number density changes with redshift, which is described by the quasar luminosity function (QLF). The QLF measures the number density of quasars per unit luminosity and is usually expressed as a double power-law, featuring a luminosity break, $L_{\star}$. This framework distinguishes two scenarios: in luminosity evolution, the luminosity varies over time while the number density stays constant, whereas in density evolution, the number density changes while the luminosity of individual quasars remains stable \cite{Richards}. Measuring the QLF is crucial for understanding the evolution of supermassive black holes (SMBHs) and for placing constraints on various cosmic backgrounds \cite{Ross}. A detailed analysis of the QLF was carried out using the SDSS-III DR9 catalog, which includes 22,301 quasars across an area of $ 2236\rm{deg}^2$. The catalog provides spectroscopic redshift measurements in the range of  $z \in [2.2, 3.5]$ and incorporates K-corrections for accurate luminosity assessments. The results confirmed the double power-law relationship in the range $z \in [0.3, 2.2]$], but also identified a clear break in this relationship around $z=2$ with $M_{I}(z=2) =-24.5$. For $z\geq 2.2$ , they found a log-linear trend in quasar luminosity amplitude and break luminosity \cite{Ross}.\\

By combining data from the SDSS and 2dFQSO surveys, researchers examined the clustering properties of quasars up to redshift $z=5$, calculating the two-point projected correlation function for quasars within $r<20 \text{h}^{-1} \text{Mpc}$. Their analysis indicated a halo mass of about $10^{13}M_{\odot}$, with a slight increase observed with redshift. Additionally, when considering the fraction of halos hosting active quasars, the typical lifetime of quasars at high redshift ($z>1$) is estimated to be around $t_{Q} \simeq 10^{8}\rm{yr}$ \cite{porci}. \\

In addition to the previously discussed reasons for selecting a redshift window between $1.2$ and $2.8$, another significant factor refers to the completeness of the quasar sample \cite{schneider}, \cite{paris}, \cite{Lyke}. The initial releases of the SDSS DRII quasar catalog, which contained only $16,713$ objects and covered an area of $1360 \text{deg}^2$ across the interval $z \in [0.08, 5.41]$, indicated the necessity for completeness in the sample. To achieve a reliable identification of quasars, robust selection criteria based on brightness, magnitude, and color, along with multi-color imaging systems, were implemented by the SDSS collaboration. This approach effectively differentiated quasars from stars and galaxies \cite{schneider}. Significant improvements were made in the SDSS DR12 quasar catalog, which incorporated photometric detections across five bands with a typical accuracy of $0.03 \text{mag}$, as well as spectroscopic measurements of redshift by analyzing strong emission lines from carbon and magnesium \cite{paris}. The deep-scan capabilities of the survey allowed to reach specific magnitude thresholds, ensuring that quasars at higher redshifts ($z>2.1$) remained detectable, leading to the discovery of $297,301$ quasars over  $9376 \text{deg}^2$. Spectroscopic follow-up was conducted after the initial photometric identification, confirming both the quasar nature and redshift measurements. The expanded survey volume increased the likelihood of finding quasars, thereby enhancing the completeness of the sample \cite{paris}. These elements and analyses collectively contribute to a better understanding of completeness and potential selection effects that may influence the observed distribution of quasars. For instance, the typical completeness value for quasars in the SDSS DR16 catalog is often around  $99.8\%$, with a contamination rate of less than  $1.3\%$ \cite{Lyke}. In summary, we started our analysis in the redshift range of $[2.2, 2.8]$ where the SDSS DR16 catalog is reasonably complete. Afterwards, we extend our study  at the lower $z$ starting at $z=1.2$. This choice was made to give us a broader perspective, as the catalog becomes increasingly incomplete for redshifts greater than $2.8$. Our main goal is to explore how the evolution of redshift unfolds in this context. This redshift range is used in other observational works. At cosmic noon, a cross-match between Gaia EDR3 and SDSS DR16 data allowed the detection of quasar pairs with angular separations between $0.3"$ to $3"$ at redshifts greater than $1.5$ (cf.  \cite{chen2} and reference therein). Additionally, the joint analysis of JWST/NIRSpec integral field unit spectroscopy, combined with Gaia's identification, led to the discovery of a supermassive black hole pair at a redshift of $z=2.17$, separated by $3.8$ kpc \cite{ishi}.

\section{2PCF for the subsample}\label{2pcf}

The gravitational processes that dictate the formation and evolution of cosmic structures are elegantly encoded in the correlation function of galaxies. Moreover, the correlation function of quasar distributions plays a crucial role, since quasars serve as exceptional tracers of the Large--scale structures, particularly at high redshifts ($z>1$) \cite{shen}, \cite{pierre}. Their unique properties allow for an exploration of the gravitational dynamics that govern the evolution of the Universe, providing valuable insights that complement analyses of galaxy clustering \cite{song}.\\

\begin{figure}[htbp]
\includegraphics[width=\linewidth]{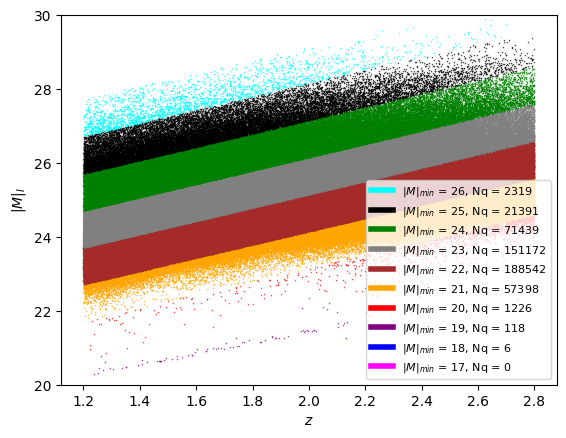}
\caption{Absolute magnitude  across the redshift range. The bands in magnitude are depicted based on the relationship $|M|_{I}(z)=|M|_{0}+ 1.8 (z-0.8)$.
\label{fig:2}} 
\end{figure}

We start our investigation with an exploratory analysis of the data drawn from the SDSS DR16 catalog \cite{Lyke}. In this endeavor, we apply a series of cuts based on the absolute magnitude in the I-band across the sample. Data are represented according to the relationship $|M|_{I}(z)=|M|_{0}+ 1.8 (z-0.8)$, where $|M|_{0} \in  [|M|_{\text{min}}, |M|_{\text{max}}] $. 
Here, the lower and upper limits of each band are defined as $|M|_{\text{min}}=18+n$, $|M|_{\text{max}}=|M|_{\text{min}}+1$, where $1 \leq n\leq 10$ (see Fig. \ref{fig:2}). This methodology enables us to discern any gradients present in the quasar data, which is crucial for the subsequent construction of the (weighted) random data set.\\

Now, we proceed with the numerical computation of the two-point correlation function (2PCF), using the capabilities of the \textit{corrfunc} package \cite{Sinha}. The \textit{corrfunc} package\footnote{\url{https://github.com/manodeep/Corrfunc.}} allows us to understand the intricacies of the correlation function, specifically by examining the autocorrelation properties and the coherence length associated with our selected subsample. By conducting this analysis, our aim is to elucidate the power-law behavior that emerges within the linear regime \cite{White}. To facilitate its exploration, we initiate the process by constructing a random catalog that mirrors the redshift interval of our original data set. The random catalog serves as a crucial reference point, enabling us to compare the observed correlations with those expected in a uniformly distributed sample.  \\

\begin{figure}[htbp]
\includegraphics[width=\linewidth]{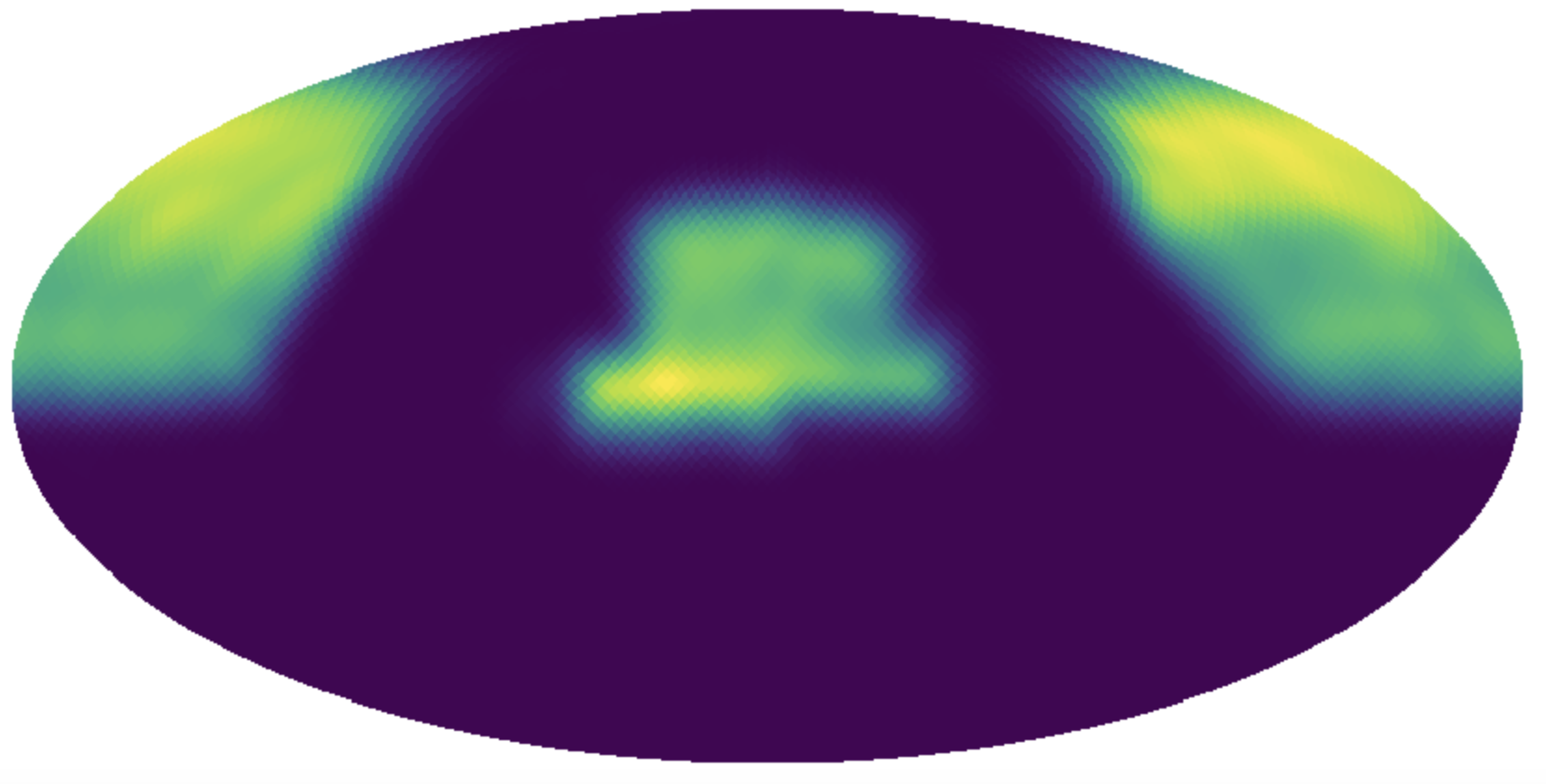}
\includegraphics[width=\linewidth]{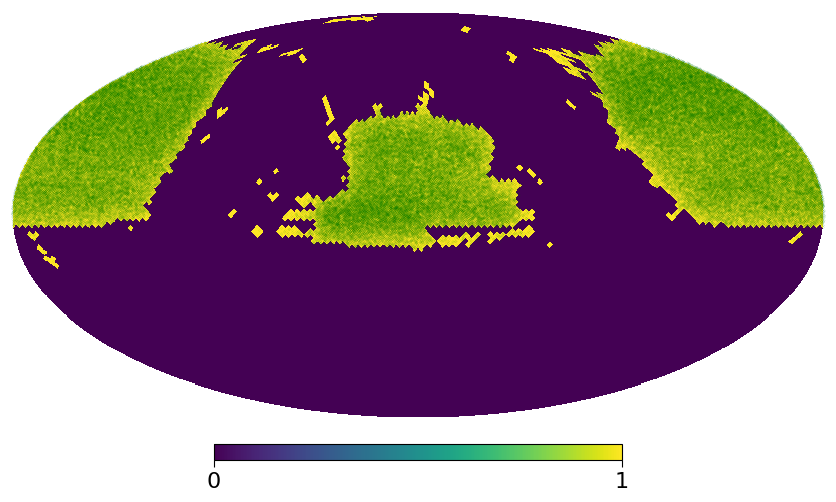}
\caption{Top Panel: We illustrate the mask employed to generate a random map at a resolution of $N_{\text{side}}=32$ incorporating a smoothing factor of $\theta_{s}=4^{\circ}$. Bottom panel: The yellow region represents the dataset points, while the green region emphasizes the random set of points produced using the probability density function approach alongside the smoothing mask. Overall, the overlap between both sets indicates a strong level of concordance
\label{fig:1}} 
\end{figure}

For a resolution of $N_{\text{side}}=32$, we construct a mask to accurately mimic the distribution of quasars across the celestial sphere. To achieve our goal, we use the \textit{Healpix} package\footnote{\url{https://healpix.sourceforge.io}} \cite{healpix}, where a $N_{\text{side}}=32$ is represented by 12288 pixels. The procedure is as follows:  We construct a smoothed and weighted map designed for quasars to populate the random catalog, which accounts for the gradient in the data \cite{Lyke}. The latter smoothed map serves as a probability density function (PDF). The process begins by generating random right ascension (RA) and declination (DEC) coordinates. We then check the compatibility of these random coordinates with the smoothed map. We also ensure that each point falls within the designated mask; thus, points are more likely to be accepted into the random catalog if they are generated in regions with a high density of quasars. We evaluated the probability of the normalized weighted map at these coordinates and compared the probability with a randomly generated filter value between 0 and 1. If the filter value exceeds the probability at the selected RA/DEC coordinate, the process is restarted from the previous step. Otherwise,  we accept and save the random RA and DEC. The key point is now that we are working directly with a map of probabilities instead of computing a PDF. In doing so, we ensure that the random RA/DEC points are concentrated in areas where quasars are more prevalent, resulting in a catalog that accurately reflects the spatial distribution of the original dataset. In Fig. (\ref{fig:1}), we present the distribution of quasars across the celestial sphere alongside its corresponding mask based on the procedure mentioned earlier.  We also verified that the distribution of the random data in redshift mirrors the trends observed in the original data sample. The comparison ensures that our randomization process preserves the underlying characteristics of the redshift distribution, allowing us to confidently assess the significance of our findings. \\

\begin{figure}[htbp]
\includegraphics[width=\linewidth]{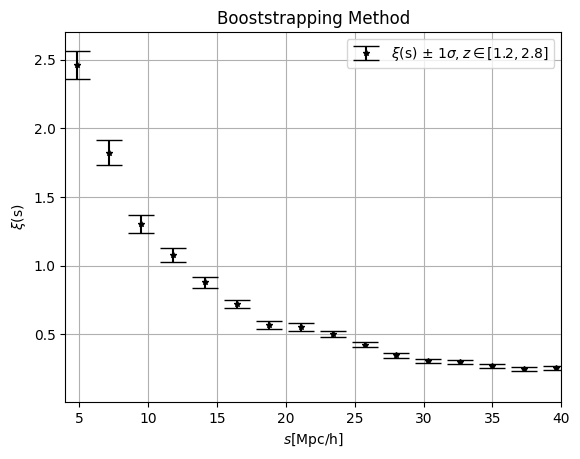}
\includegraphics[width=\linewidth]{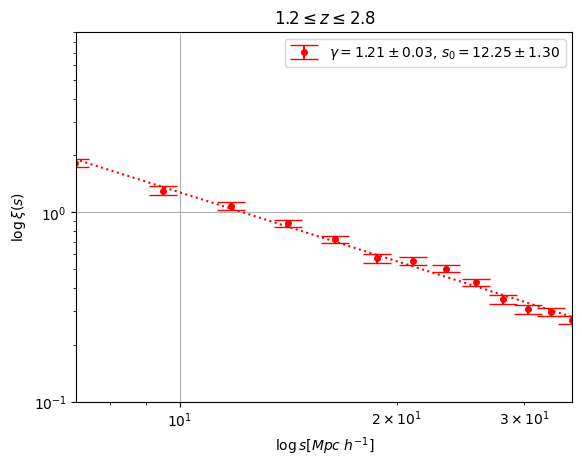}
\caption{Top panel: Two-point correlation function obtained using the LS
estimator for the total sample, $z\in [1.2, 2.8]$. Red point represents the correlation function with its error bars. Bottom panel:  The dashed-red curve indicate a best linear fit in the logarithm scale. In the linear regime, the correlation function can be parametrized as $\xi(s)=(s/s_{0})^{-\gamma}$ with $\gamma=1.21\pm 0.03$ and $s_{0}=12.25\pm1.30$.\label{fig:3}} 
\end{figure}

To compute the correlation function in the redshift space  using the \textit{corrfunc} package \cite{Sinha}, we make use of the \cite{LS} estimator provided it reaches minimal variance, 
\begin{equation}
\label{lsesti1}
\xi(s)=\frac{DD -2DR +RR}{RR},
\end{equation}
where $DD$ the number of distinct data pairs, $RR$ the number of different random pairs, while $DR$ is the number of cross-pairs between the real and random catalogs within the same distance bin. In addition, the \cite{LS} estimator proves advantageous, as it effectively manages edge effects, allowing the appropriate consideration of missing data beyond the sampled region, which may possess rather irregular boundaries. In Fig. (\ref{fig:3}), we show the 2PCF obtained with the random set constructed before.  \\

We applied the Bootstrap method to estimate errors directly from the calculated correlation function \cite{nor}, \cite{perci}. Specifically, we divide the survey into $N_{s}$ subsamples and perform a random resampling with replacement to create bootstrap samples $N_{s}$. For each bootstrap sample, we compute the correlation function based on the resampled data.  The covariance matrix $C_{ij}$  between two estimates $\xi_{i}$ and $\xi_{j}$ at different bin values $s_{i}$ and $s_{j}$ is given by  \cite{nor}, \cite{perci}:
\begin{equation}
\label{lsesti2}
C_{ij}=\frac{N_{s} -1}{N_{s}} \sum^{N_{s}}_{k=1}\big(\xi^{(k)}_{i}-\langle\xi_{i}\rangle\big)
\big(\xi^{(k)}_{j}-\langle\xi_{j}\rangle\big),
\end{equation}
where $\xi^{(k)}_{i}$ is the estimate of the correlation function at the bin $s_{i}$ from the $k$-th bootstrap sample, while $\langle\xi_{j}\rangle$ stands for the mean of the estimates across all the bootstrap samples, specifically $\langle\xi_{j}\rangle=N^{-1}_{s}\sum^{N_{s}}_{k=1} \xi^{(k)}_{j}$. In the latter analysis, we consider $N_{s}=100$, and the number of bins in the $s-$variable is 60. We also performed a linear regression analysis using the estimated correlation function along with its corresponding bootstrap errors. We require that the convergence condition $\sigma^{2}_{\xi}/\langle \xi \rangle ^{2} \simeq (0.1-0.01)$ hold.  In the linear regime, the correlation function can be expressed as a power law, given by $\xi=(s/s_{0})^{-\gamma}$. Our linear fitting results in an exponent index of $\gamma=1.21\pm 0.03$ and a correlation length in the redshift space of $s_{0}=(12.25 \pm 1.30)\text{Mpc/h}$ for a determination coefficient of $R^{2}\simeq 0.98$. The values obtained are in agreement with the existing literature on quasars with intermediate redshifts \cite{White}. It is important to note that the correlation maintains its physical significance for $s<30 \text{Mpc}h^{-1}$, after which it tends to diminish toward zero.  Our analysis demonstrates that the correlation functions for the different magnitude subsets remain largely consistent on the values of $\gamma$ and $s_{0}$.

\section{Quasar Pair Catalog}\label{QPC}

We outline the primary criteria used to identify quasar pairs within the subsample of quasars at redshifts $1.2 \leq z \leq 2.8$ extracted from the SDSS DR16 catalog \cite{Lyke}. The process of identifying these quasar pairs through spectroscopic measurements requires a series of physically reasonable assumptions. To be consistent with the limitation given by the SDSS DR16 sample\footnote{The quasar sample from SDSS DR16Q exhibits characteristic statistical redshift uncertainties on the order of $|\Delta z| \simeq 0.001$ \cite{Lyke}.}, we propose that a pair of quasars must have the following features:
\begin{itemize}
\item The quasar pairs must adhere to a limiting velocity constraint such that $|\Delta V|=c(z_{2}-z_{1})\leq 900 \rm{km}/s$. 

\item  The quasars that make up the pairs should be in close proximity, specifically with a projected distance of $d_{p}\leq 2 \rm{Mpc}$.
\end{itemize}
These two conditions together define the cylindrical volume within which the quasar pairs are located. In addition, it is expected that within any given pair of quasars one will exhibit a higher luminosity, reflected by a smaller (and thus more negative) value of apparent magnitude, $m_{i}$. The method developed to identify the aforementioned quasar pairs can be summarized as follows. Each quasar in the catalog is assigned a unique ID number along with all its physical properties, including angular coordinates (RA, DEC), redshift, apparent magnitude, and luminosity expressed in solar units. We use the Planck 2018 background as a fiducial cosmology for numerical computations with the \textit{Astropy} package\footnote{\url{https://www.astropy.org/}} \cite{astropy}. The angular diameter distance for each quasar is calculated on the basis of its redshift and the Planck 2018 cosmology. The subtended angle in radians is determined from a specified maximum distance and is subsequently converted to degrees. The RA and DEC coordinates are translated into \textit{Healpix/Healpy}\footnote{\url{https://healpy.readthedocs.io/en/latest/}} pixel indices for spatial analysis \cite{healpix}. Each pixel corresponds to a specific location in the celestial sphere, and we employ a high resolution of $N_{\text{side}}=2048$. We store the vectors for each pixel for future use in the \textit{query-disk} routine and create a list to keep track of neighboring pixels. For each quasar, neighboring pixels within a defined angular distance are identified, facilitating the analysis of local quasar environments. The radius in radians is computed using the angular distance, and the neighboring pixels within the specified disk are recorded. The algorithm then searches for nearby quasar pairs based on their redshift and angular distance, ensuring that no pair is identical and that they lie within a predetermined redshift tolerance. Filter out duplicate pairs employing ordered tuples and retains only the closest pairs, constrained by a physical distance criterion of $d_{p}\leq 2\text{Mpc}$. \\
\begin{figure}[htbp]
\includegraphics[width=\linewidth]{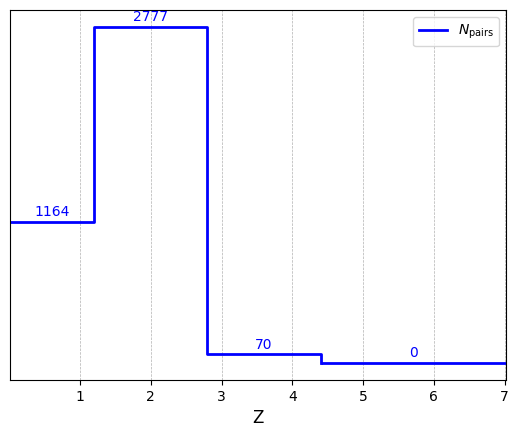}
\includegraphics[width=\linewidth]{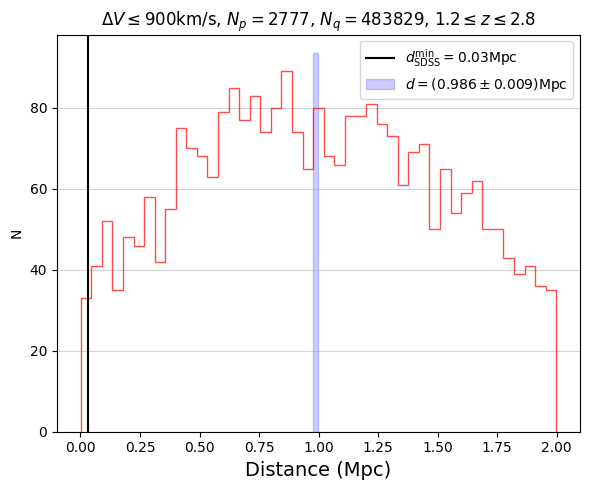}
\caption{Top panel: Distribution of the number of quasar pairs across the complete redshift window of the SDSS DR16 quasar catalog. The highest concentration of pairs is found within the redshift range $z \in [1.2, 2.8]$, totaling 2777 pairs. Overall, the catalog contains a maximum of 4011 pairs.
Bottom panel: The distribution of quasar pairs detected within a cylindrical volume characterized by a projected radius of $d_{p}\leq 2\text{Mpc}$ and a length of $|\Delta V|= 900 \text{km}/s$ is presented. The black vertical lines denotes $d^{\text{min}}_{p}\simeq 0.03 \text{Mpc}$, representing the minimum projected separation at which a quasar pair can be reliably identified.  This threshold is determined by the angular resolution of the SDSS camera, set at $\theta_{SDSS} = 5^{"}$  along with a mean redshift value of 
 $z_{\text{mean}}=2.4$.
\label{fig:4}} 
\end{figure}

The distribution of the identified quasar pairs is shown in Fig. (\ref{fig:4}). In the SDSS DR16 catalog, with redshifts ranging from $z = 0.0007$ to $z = 7.023$, a total of 4011 quasar pairs have been identified. Most of these quasar pairs are found in the first two redshift bins: $z \in [0.0007, 1.2]$ and $z \in [1.2, 2.8]$ with pairs detected beyond $z = 4.4$. The highest detection rate  occurs in the second redshift interval $z \in [1.2, 2.8]$, accounting for a sample of 2777 pairs where we will focus our study. For example, the histogram of projected distances in Mpc for all 2777 quasar pairs is also shown in Fig. (\ref{fig:4}). The average distance between pairs is  $d=(0.986 \pm 0.5009) \text{ Mpc}$ with only $21$ pairs have projected separations below the minimum projected distance resolved by the SDSS, $d^{\text{min}}_{p}\simeq 0.03 \text{Mpc}$. Overall, the peak of the histogram suggests that quasar pairs tend to cluster around this average projected distance.\\

\begin{figure}[htbp]
\includegraphics[width=\linewidth]{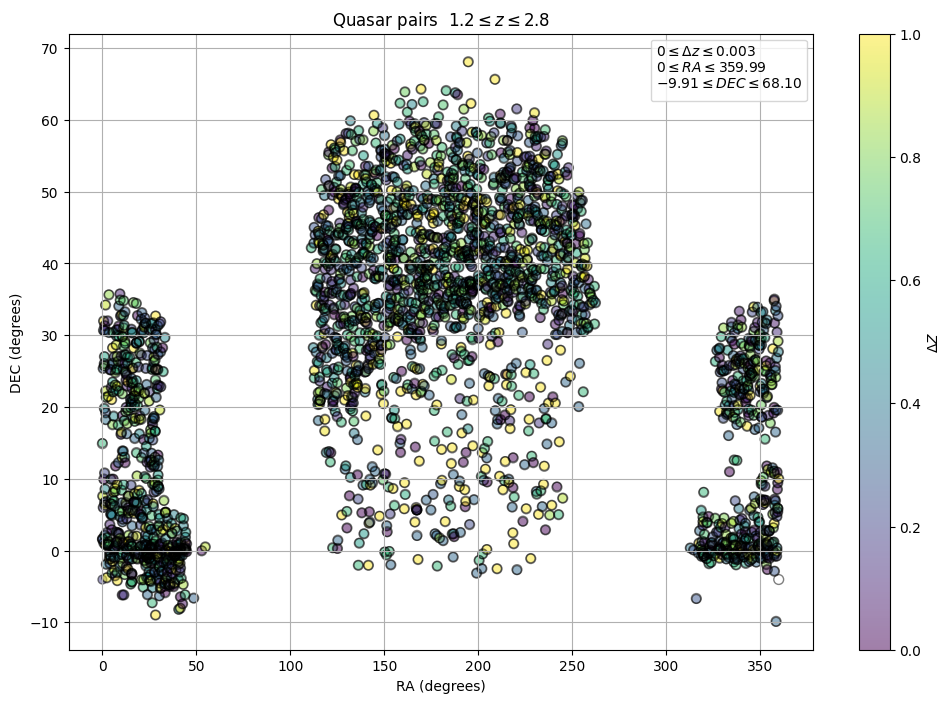}
\caption{For the subset of 2777 quasar pairs that meet the criterion $|\Delta V|\leq 900 \text{km}/s$, a scatter plot illustrating these pairs is shown. This visualization incorporates a normalized color bar that conveys the variation of $\Delta z$  among the quasar pairs, effectively representing this continuous variable through a color scale.
\label{QD}} 
\end{figure}

We also examine various bounds of $|\Delta V|$ to investigate how the number of pairs detected changes as this condition is relaxed within the redshift interval $z \in [1.2, 2.8]$. For the condition $|\Delta V|\leq 600 \text{km}/s$, the number of pairs detected decreases to 1915, representing $39\%$ of the total. In contrast, when we consider $|\Delta V|\leq 1200 \text{km}/s$, the number of pairs detected increases to 3605, which represents $73\%$ of the entire subsample. Consequently, we will focus on the most conservative case moving forward, which corresponds to $|\Delta V|\leq 900 \text{km}/s$.  The distribution of quasar pairs within the RA-DEC plane is illustrated in Fig. (\ref{QD}) for the conservative scenario.\\

We address another observational  point regarding the detection of quasar pairs. Firstly, we pose a fundamental question: Is the SDSS camera capable of effectively resolving two quasars that constitute a confirmed pair? This query is essential for understanding the limitations of observational data and the ability of the SDSS to distinguish quasar pairs, which is crucial for accurate pair identification and subsequent analysis. To answer that question, we use a conservative angular resolution for the SDSS telescope \cite{Lyke}, with $\theta\geq\theta_{SDSS} = 5^{''}$, and a mean redshift value of $z_{\text{mean}}=2.4$. The physical distance corresponding to the SDSS angular resolution yields approximately $d^{\text{min}}_{p}\simeq 0.03 \text{Mpc}$ [cf. Fig. (\ref{fig:4})]. Such findings are significant in light of previous observations, which indicated that the distribution of distances between quasar pairs is around $d=0.986\text{Mpc}$. Thus, we can conclude that the SDSS camera is capable of resolving these objects. \\

Another important question is whether the quasar pair sample might be contaminated. For example, background galaxies can create multiple images of a single quasar, making it difficult to identify true quasar pairs. To tackle this issue, we can reassess the quasar pair sample by focusing only on pairs with $\Delta V\leq \sigma^{SDSS}_{V}$, where  $\sigma^{SDSS}_{V}(z)=60\rm{km}/s$ represents the error in the spectroscopic measurements from the SDSS DR16 catalog \cite{Lyke}, and $\Delta V=c(z_{2}-z_{1})$.Our analysis revealed that out of the 2777 detected quasar pairs, only $9.9\%$ could be potentially contaminated. To investigate these cases further, we used the specific coordinates of these quasars to examine their corresponding spectra from the SDSS DR16 catalog. However, it's important to note that the SDSS DR16Q catalog has already looked into several contamination issues within the full sample, including duplicity due to the observational strategy, reporting up to $1.3\%$ of potential contamination \cite{Lyke}. Our criteria for assessing contamination are as follows:
If both quasars in a pair show identical features in their spectra, it suggests they may be different images of the same quasar, influenced by foreground galaxies. On the other hand, if the elements of a quasar pair exhibit different amplitudes and positions of their emission lines, we can confidently conclude they are not multiple images of the same object. The position of each object in the sky and its spectrum can be accessed through the \textit{skyserver-dr16 wesbsite} \footnote{\url{https://skyserver.sdss.org/dr16/en/tools/chart} }.


Using the sky-server DR16, we obtained the spectra associated with each quasar by inputting their angular coordinates on the sky. After normalizing each spectrum to its maximum value, we set a criterion: if the spectra of quasar pairs differed by less than $5\%$, they could be considered as a single quasar producing multiple images, especially if the redshifts of the component quasars were very close. To analyze the spectral lines and their ratios, we employed a package called \textit{PyQSOFit} \cite{gu}. For pairs where $0<\Delta V\leq 60 \text{km}/s$, we determined that only $30\%$ could be the same lensed object. For our total sample of 2777 pairs, we expect a contamination lesser than $3.5\%$ due to lensing effects, which amounts to 66 cases. A similar approach has been used to create a new quasar pair catalog based on the DESI DR1 quasar sample leading to the identification of $1,842$ new pairs by applying a projected separation criterion of  $d_{p}\leq 110\text{kpc}$ and a radial velocity difference limit of $\Delta V\leq 600 \text{km/s}$. This method not only improves the completeness of the quasar pair sample but also helps ensure a more accurate identification of real pairs, reducing the potential contamination from background sources \cite{J}.\\

To further characterize the quasar pair catalog, we perform a numerical estimation of the comoving number density of quasars and quasar pairs within the redshift interval of $[1.2, 2.8]$. The fraction of the sky surveyed is approximately $0.65\%$ covering $26732\text{deg}^2$, leading to a numerical (comoving) density of $n_{q}\simeq 5.8 \times 10^{-8}\text{Mpc}^{-3}$. This value aligns well with the findings reported in \cite{cro}; namely, $n_{q}\simeq 8.2 \times 10^{-9}\text{Mpc}^{-3}$. Taking into consideration a minor sky fraction for the quasar pairs $f_{\text{sky, pair}}=0.54$, we observe a notable reduction in the estimated numerical (comoving) density of these pairs. It approximately yields $n_{\text{pair}}\simeq 2.4 \times 10^{-10}\text{Mpc}^{-3}$. The latter finding highlights the significant influence that observing a restricted area of the sky, combined with a relatively small sample of detected pairs, has on the estimated density of quasar pairs, particularly in contrast to the densities associated with individual quasars \cite{cro}.


\section{Luminosities and Color-Color Maps}\label{Lum}
 In this section we explore wether quasar pairs exhibit particular properties with respect to quasars in isolation. Here we focus on the luminosity of the quasar pairs \cite{lum2}, as well as the luminosity ratio of the pair members.\\

\begin{figure}[htbp]
\includegraphics[width=\linewidth]{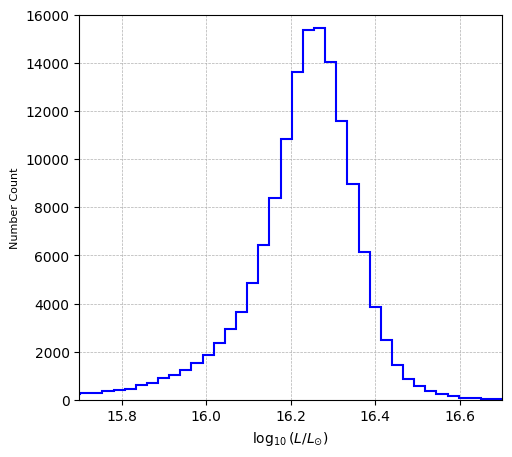}
\caption{Histogram of the quasar luminosity distribution after applying the K-corrections. The horizontal axis is displayed on a logarithmic scale.\label{fig:7}} 
\end{figure}

Quasar luminosties must take into account the K-correction, \cite{Ross}. Our approach follows  \cite{Lyke}, using a pre-calibrated K-correction table from \cite{Richards}. The K corrected absolute magnitude in the I-band is calculated using the relations:
\begin{equation}
\label{bol1}
m_{I,\text{ap}} = M_{I,\text{abs}} + DM(z) + K(z) +A(z),
\end{equation}
where $m_{I,\text{ap}}$ is the apparent magnitude, $M_{I,\text{abs}}$ is the absolute magnitude in the I-band, $DM(z)=\log_{10}\left[D_{L}(z)/10 \rm{pc}\right]$ is the distance modulus, and $A(z)$ represents extinction. The K-correction, found in the second-to-last term of the equation, can be specified at $z=2$ as detailed in Appendix B of \cite{Ross}. Our method for integrating K-correction closely follows the procedures from \cite{Ross} and \cite{Lyke}. We use the Planck 2018 fiducial cosmology for distance calculations. The absolute magnitude $M_{I,\text{abs}}$ for a given quasar is obtained by correcting the apparent magnitude for extinction, distance modulus, and K-correction. Apparent magnitudes in the I-band are taken from the \textit{PSFMAG} column in the FITS file of the SDSS DR16 QSO catalog \cite{Lyke}\footnote{\url{https://www.sdss4.org/dr16/algorithms/qso_catalog}}. K-corrections come from Richards' table, which is limited to a specific redshift range \cite{Richards}. For each redshift $z$, we find the closest entry in the K-correction table to extract the corresponding K-correction value. This process allows us to calculate the corrected absolute magnitude  $M_{I,\text{abs}}$ for each quasar. Once we have $M_{I,\text{abs}}$, we can convert it to  luminosity in the I-band in solar units using:
\begin{equation}
\label{bolr}
L=10^{-0.4(M_{I,\text{abs}}-M_{I,\odot})}L_{\odot},
\end{equation}
where $L_{\odot}$ is the solar luminosity and $M_{I,\odot}$ is the absolute magnitude of the Sun in the I-band. Thus, we obtain reliable estimates of the  luminosities in the I-band allowing for further analysis of quasar in pairs and in isolation. In Fig. (\ref{fig:7}), we display the distribution of luminosities, showing that $L=(1.75\pm 0.52)\times 10^{16} L_{\odot}$.\\


It is now well understood that galaxy evolution, shaped by AGN feedback, often leads to quasars residing in low-mass dark matter halos \cite{fani}. Furthermore, the number density of quasars per unit luminosity follows a double power-law relationship within the redshift range $z \in [0.3,2.2]$, while showing log-linear behavior for $z \geq 2.2$ \cite{Ross}. This suggests that quasar pairs-acting as proxies for protoclusters \cite{muldrew}, \cite{noble}-may share similarities due to their common dark matter halo. \\

Here we aim at addressing if quasar pairs within the same dark matter halo represent a distinct class of objects, rather than just a simple combination of two individual quasars from the sample. To address the latter point, we  examine the distribution of luminosity ratios, $L_{1}/L_{2}$, for the quasar pairs, where $L_{1}$ is the smaller luminosity and $L_{2}$ is the larger one. This analysis will allow us to investigate the intrinsic relationships between the luminosities of these unique objects, offering insights into their relative behaviors. Additionally, we will generate 100 random sets of control pairs derived from the isolated quasar set. These ensembles will not satisfy the typical constraints of $\Delta V\leq 900 \text{km}/s$ and $d_{p}\leq 2 \text{Mpc}$ \footnote{A similar analysis was carried out on 500 random sets of synthetic pairs generated from the isolated set, showing no significant deviations.}. This approach will provide a baseline for comparison, which is essential for understanding the characteristics of the quasar pairs and their luminosity distributions in relation to the control pairs. By contrasting the observational data with these control constructs, we aim to identify any significant patterns or anomalies that may arise, thereby enriching our comprehension of these astrophysical entities.\\  
 
\begin{figure}[htbp]
\includegraphics[width=\linewidth]{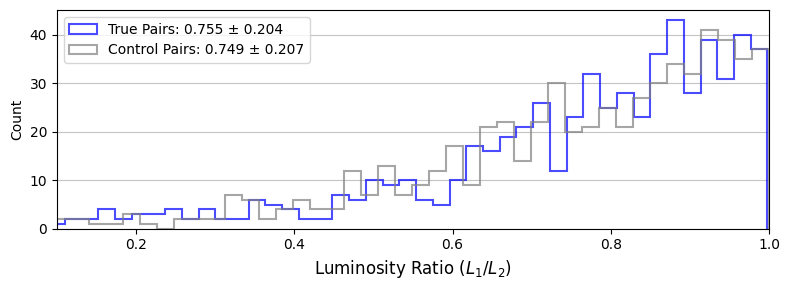}
\caption{Distribution of the luminosity ratios for true pairs and control ones. The true quasar pairs are indicated in blue, whereas the control pairs are shown in gray.\label{fig:8}} 
\end{figure}

We construct the control pair sample imposing that the angular separation between the two objects exceed $5.7^{\circ}$ (1 radian), and the maximum allowable redshift difference is restricted to $\Delta z \leq 0.003$.  We also verified that the histograms of $\Delta z$ for the true pairs and control pairs are quite similar. This consistency in their distribution assures that there is no significant difference in the redshift differences between the two groups, indicating that the selection criteria for both true and control pairs are comparable. We generated 500 sets of control pairs, each containing the same number of detected pairs, resulting in 615 control pairs per set. We focused on assembling the new set using a random algorithm to avoid any selection bias \footnote{While our focus is on the redshift interval $z \in [2.2, 2.8]$, it is noteworthy that a similar analysis has been conducted across other redshift bins, yielding comparable results.}.  \\

Fig. (\ref{fig:8})  illustrates the differences between the true quasar pairs and the control pairs. The mean luminosity ratio $L_{1}/L_{2}$ shows  significant variations between these two datasets. 
When we calculate the fraction of real pairs that meet the condition $F_{\text{true pair}}(L_{1}/L_{2}>0.75)$ and compare it to the fraction for the control pairs $F_{\text{control pair}}(L_{1}/L_{2}>0.75)$, we can determine the ratio of these fractions, defined as ${\cal{R}}=F_{\text{true pair}}/F_{\text{control pair}}$. The goal of calculating $F_{\text{control pair}}(L_{1}/L_{2}>0.75)$ is to see if the control pairs can mimic the histogram of quasar pairs that show $L_{1}/L_{2}>0.75$.\\

The results show that  ${\cal{R}}>1$, highlighting that the two sets of pairs have fundamentally different properties, likely due to the different accretion mechanisms affecting the luminosity ratios $L_{1}/L_{2}$. Our analysis finds that $62.32\%$ of true pairs exhibit $L_{1}/L_{2}>0.75$, while only $57.56\%$ of control pairs meet this criterion. Thus, the calculated value of ${\cal{R}}=1.08$ indicates that true pairs show a luminosity ratio $L_{1}/L_{2}>0.75$ compared to their control sample counterpart. Moreover, our findings are consistent across all 500 datasets of control samples, each containing 615 pairs, reinforcing the robustness of our conclusions.\\

\begin{figure}[htbp]
\includegraphics[width=\linewidth]{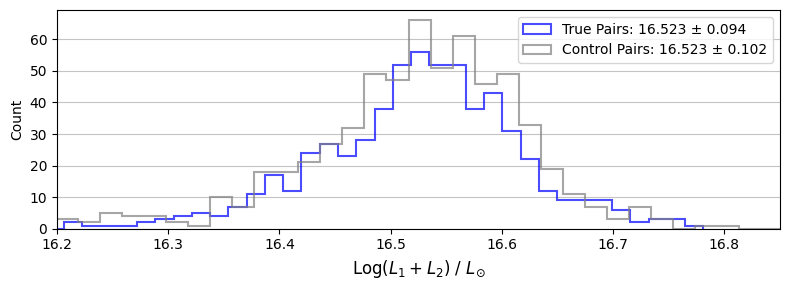}
\caption{Distribution of the logarithm of the sum of luminosities for true pairs and control pairs.
The true quasar pairs are indicated in blue, whereas the control pairs are shown in gray.\label{fig:9}} 
\end{figure}

Another approach to investigate the validity of random-generated data sets compared to true pairs is to analyze the behavior of the total luminosity of the pair, i.e., the sum of the member luminosities. This is motivated by the possibility of a different behavior for quasars in association with respect to randomly selected objects (notice that the results of the luminosity ratio analysis could imply both quasars to have either low or high luminosities). 
As it can be seen in Fig. (\ref{fig:9}) the resulting distribution of the logarithm of the sum of the luminosities differs for true pairs and control pairs. The histogram of the true pairs shows a narrower spread around its mean, indicating that these pairs are more closely clustered in terms of their luminosity ratios. In contrast, the histogram for the control pairs exhibits a broader spread, indicating a larger luminosity variability among them. This difference in distribution reinforces the notion that true pairs are more consistent in their luminosity characteristics in contrast to the control sample pairs tend which  display a wider range of luminosity ratios.\\


Our analysis shows a significant difference between the two datasets, suggesting that the mechanisms driving the luminosity characteristics of quasars in association are intrinsically distinct. These differences may largely arise from the co-evolution of quasars with their surrounding environments, which influences their accretion processes and, in turn, their luminosity profiles \cite{porci}. Such interactions with the environment can significantly shape the properties of quasars, leading to observable variations in luminosity distributions \cite{lum}.\\

\begin{figure}[htbp]
\includegraphics[width=\linewidth]{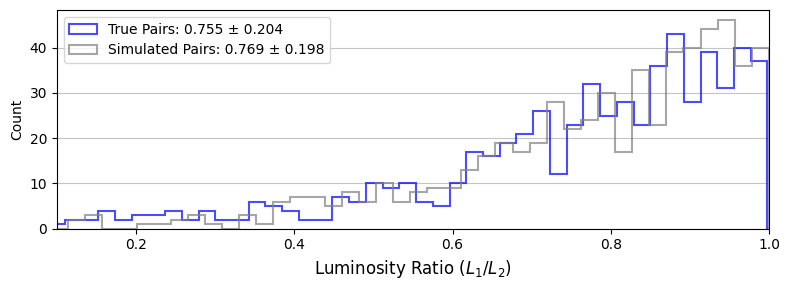}
\caption{The distribution of  luminosity ratios for true pairs and simulated pairs is illustrated. The true quasar pairs are represented in blue, while the simulated pairs are depicted in gray.
\label{fig:9x}} 
\end{figure}

We aim to further quantify the differences observed between true quasar pairs and their control counterparts. The main goal is to evaluate whether a modified set of pairs from the control sample, hereafter named "simulated" quasar pairs, can effectively approximate the real data. In this approach, we focus on the tercile of the control sample associated with the less luminous quasar and increase its luminosity by $30\%$.\\

Figure (\ref{fig:9x}) shows that this procedure creates a simulated quasar pair sample with a distribution of luminosity ratios  $L_{1} /L_{2}$ that closely resembles the actual data, showing the strength of  similarity effect. 
Fundamentally, this observation indicates that true pairs exhibit closely similar luminosity values. In contrast, control pairs show a more pronounced disparity between their most and least luminous members. This distinction may imply that 
reinforcing the hypothesis of a common cosmic evolution of our pair sample.\\

\begin{figure}[htbp]
\includegraphics[width=\linewidth]{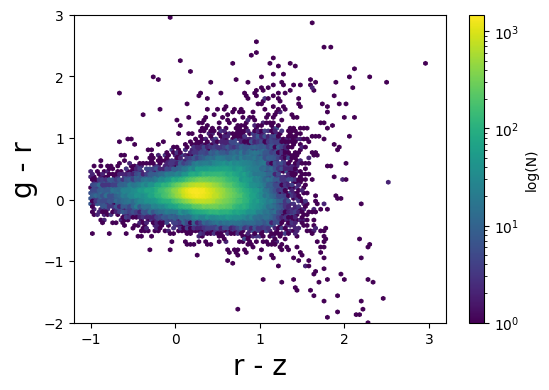}
\includegraphics[width=\linewidth]{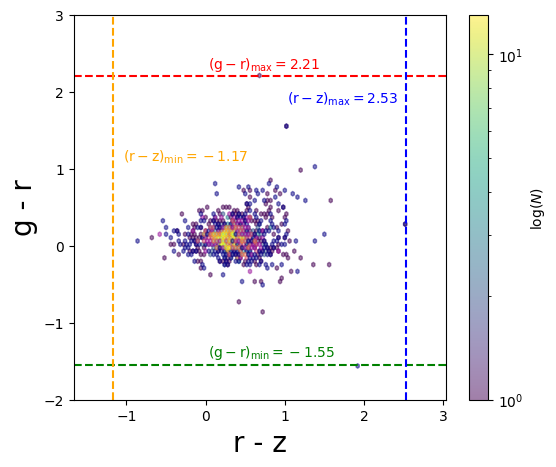}
\caption{Top panel: Color-color map for the entire subsample of quasars. Bottom panel: Color-color map for the detected quasar pairs. Here,  $\log(N)$ represents the logarithm of the number of quasar pairs that fall within each region of the color-color map. Notice that the maximum or minimum values may be outside the range of displayed data points in some cases. \label{cmap}} 
\end{figure}

\begin{figure}[htbp]
\includegraphics[width=0.9\linewidth]{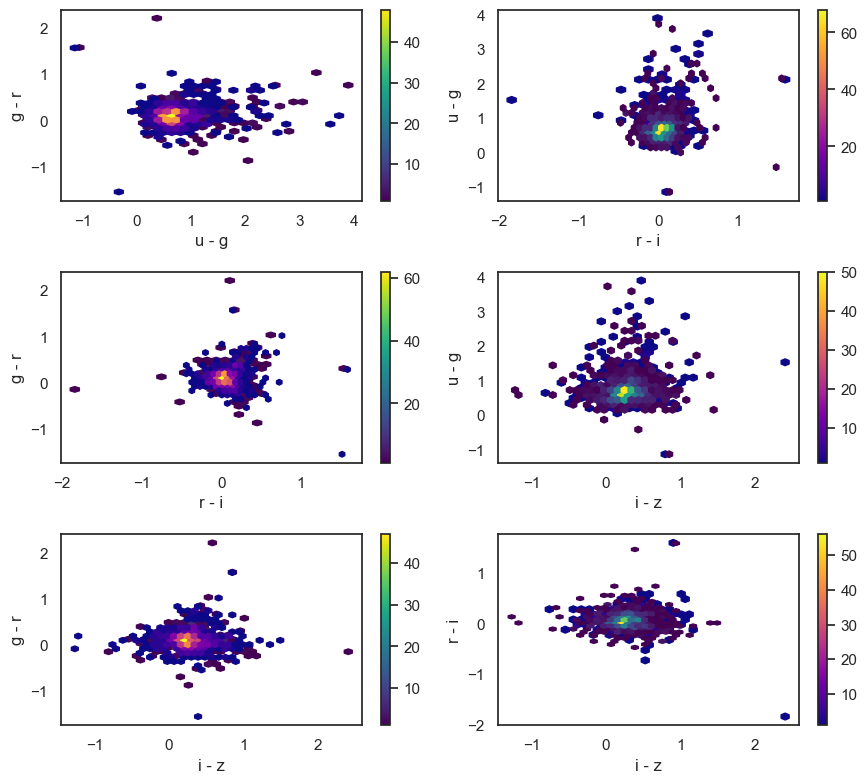}
\caption{Different color-color maps for the quasar pairs are shown. \label{color}} 
\end{figure}

In the final stage of characterizing the quasar pairs, we perform a comparative analysis of the color-color map of these pairs against that of the entire selected sample. As illustrated in Fig. (\ref{cmap}), the color-color map derived from the PSF magnitudes in the $g$, $i$, $r$ and $z$ bands for the complete subsample reveals a range of values given by $-6.04\leq (g-r)\leq 5.64$ and $-3.68\leq (r-z)\leq 7.15~$  \cite{Lyke}. However, when we focus on the detected pairs, these maximum and minimum values are significantly constrained, $-1.55\leq (g-r)\leq 2.21$ and $-1.17\leq (r-z)\leq 2.53$. It is important to note that $90\%$ of the data points associated with the quasar pairs fall within the interval $(r-z)\times (g-r)\in [-0.10, 0.85]\times[-0.17, 0.42]$. The latter results indicate older stellar populations or dust effects along with the possibility of bluer colors signifying ongoing star formation. This is consistent with the current understanding of quasar environments and their host galaxies \cite{KH2}.  For the $90\%$ of the detected quasar pairs, we obtain $(i-z)\times (r-i)\in [-0.06, 0.69]\times[-0.13, 0.30]$. Once again, the negative minimum values suggest the presence of younger, hotter stars, while the positive maximum values indicate the influence of older stars or dust obscuration. These interpretations are well supported within the framework of quasar studies \cite{Richards}. Different sequences of color-color maps are displayed in Fig. (\ref{color})

\section{Cross-correlation Quasar-Pair}\label{cross}
We intend to broaden our investigation by inspecting the cross-correlation between quasar pairs and individual quasars.  The rationale behind examining the cross-correlation lies in the fact that, within a defined redshift range, quasars act as tracers of mass distribution, albeit with an inherent bias. Consequently, our aim is to analyze the clustering behavior of quasar pairs in relation to the underlying background represented by the quasar sample that is not included in these pairs. Such an approach will deepen our understanding of the Large--scale structure of the Universe and refine our mapping of the cosmic web, especially as these entities engage and coexist with galaxies in the vast Universe \cite{menard}. \\

To ensure that the identified pairs are indeed physical composite objects located within the same dark matter halo and are likely experiencing similar accretion processes—rather than being merely coincidental nearby quasars—we will implement a series of verification tests. Among these, we will conduct an analysis of the cross-correlation of signals between the quasar sample and the corresponding subsample of pairs, as we stated before. We implement the same protocol as previously done. Using the \textit{corrfunc} package \cite{Sinha}, we calculate $\xi(s, \mu)$, subsequently converting this to the number of pairs with the assistance of the \textit{convert-3d-counts-to-cf} routine. That enables us to derive the \cite{LS} estimator for the cross-correlation,
\begin{equation}
\label{cc2}
\xi^{pq}(s,\mu)=\frac{D_{q}D_{p}-D_{q}R_{p}-D_{p}R_{q}+R_{q}R_{p}}{R_{p}R_{q}},
\end{equation}
where $D_{q}$ represents the set of quasars belonging to the subsample, and $D_{p}$ denotes the dataset of the quasar pairs, using the mid-vector between the two quasars to represent each pair. We implement a method to generate an associated random set that satisfies the specifications outlined in the \textit{corrfunc} manual \cite{corfun}. Specifically, we ensure that the size of the random set and the data set is consistent with the relationship $N(R_{p})=3N(D_{p})$. The latter approach improves statistical robustness and mitigates any potential biases in our correlation measurements. We integrate the previous method with a bootstrap algorithm \cite{perci}. By applying the bootstrap technique, we can derive confidence intervals and improve the robustness of our cross-correlation analysis. This iterative process allows us to quantify the uncertainties associated with our measurements, ultimately enhancing the statistical significance of our findings. We consider one bin in the $\mu$-variable with $\mu_{\text{max}}=1$ \cite{corfun} and 60 bins in the separation $s$-variable.  The resampling number employed in the bootstrap method \footnote{It should be emphasize that  the bootstrap algorithm offers several advantages in our statistical analysis. To be more precise, it provides a more robust estimate of uncertainties by allowing replacement resampling, which better captures the underlying distribution of the data. It is particularly beneficial when the quasar pair dataset size is small compared to the larger quasar sample, ensuring that we account for variability and improve our confidence in the derived results.} is set to $N_{\text{rs}}=100$ \footnote{We investigated various values of the resampling number $N_{\text{rs}}$ and found that the results remain largely invariant, showing no significant variation across the different configurations tested.}, consistent with the approach outlined in \cite{perci}. \\
\begin{figure}[htbp]
\includegraphics[width=\linewidth]{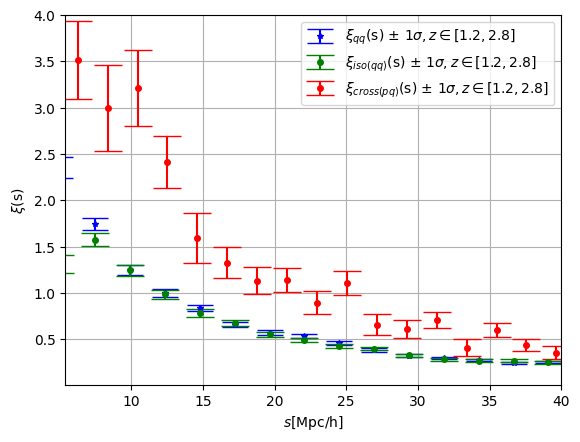}
\includegraphics[width=\linewidth]{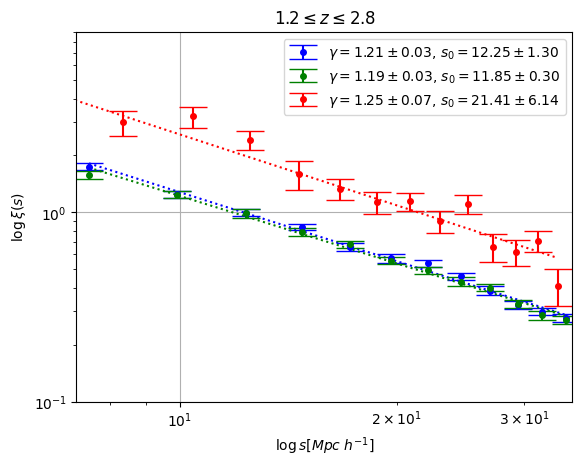}
\caption{Top panel: The comparison of the two-point cross-correlation function, derived using the LS estimator for the quasar pair-quasar within the $1.2 \leq z\leq 2.8$ redshift window, is presented alongside the auto-correlation for the full sample in that interval and the auto-correlation of the isolated quasars. Bottom panel: The linear regressions for the three aforementioned cases are depicted using a logarithmic scale to facilitate the visualization of their slopes. The estimation of  $(\gamma, s_{0})$ remains consistent across all three cases. \label{fig:11}} 
\end{figure}

To begin, we aim to illustrate the overall behavior of the cross-correlation amplitude across the entire redshift range of interest,  $1.2 \leq z\leq 2.8$. Subsequently, we will revisit this cross-correlation function under various scenarios, which will involve adjusting the redshift bins and exploring different combinations of cuts based on $L_1/L_{2}$ and  $\log_{10}\big(L_{1} + L_{2}\big)/L_{\odot}$. The aforementioned cuts are well-founded, as they are supported by the luminosity analysis of quasar pairs presented in section \ref{Lum}.\\

 In Figure (\ref{fig:11}), we present the cross-correlation function for this specified interval, comparing its signal to both the autocorrelation of all quasars and that of isolated quasars—those not associated with any pairs. Our analysis indicates that the cross-correlation between the pairs and isolated quasars does not produce a stronger signal than the overall cross-correlation derived from the complete sample within this interval. We observe that the cross-correlation signal between pairs and quasars consistently surpasses the autocorrelation signal of quasars by a significant margin. The signal is clearly stronger than the auto-correlation for separations of $s<25 \text{Mpc}\text{h}^{-1}$. While both correlation signals exhibit a similar decline at larger separations, as illustrated in the log-log plot in Fig. (\ref{fig:11}), the correlation scale for quasar pair correlations is approximately $s_{0,pq}\simeq 21.41$, nearly doubling the correlation length of the quasar sample, which is  $s_{0,qq}\simeq 12.25$. The latter observation indicates that quasar pairs within this range demonstrate significant clustering, implying a stronger spatial association influenced by underlying physical processes. Such clustering is one of the reasons why quasar pairs can be regarded as proxies for protoclusters \cite{muldrew}, \cite{noble}. \\

\begin{figure}[htbp]
\includegraphics[width=\linewidth]{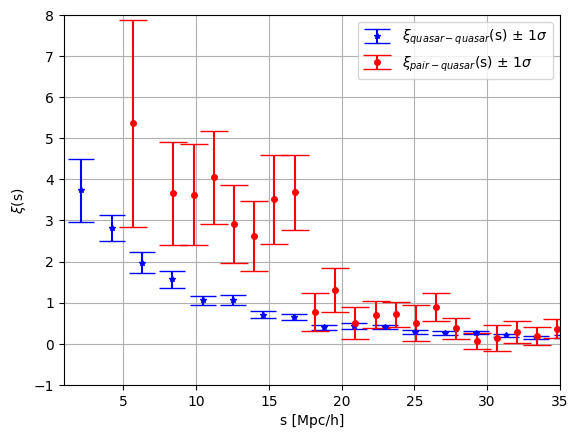}
\includegraphics[width=\linewidth]{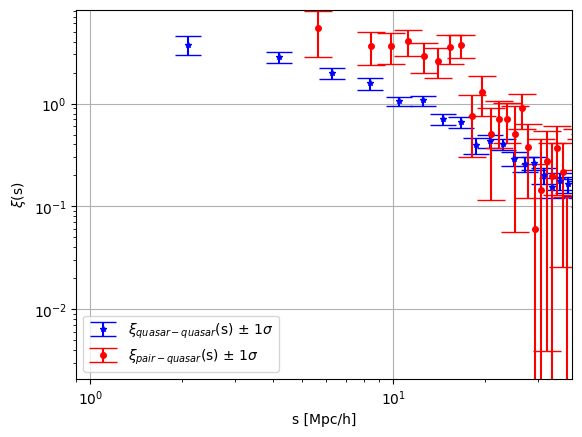}
\caption{Top panel: The two-point cross-correlation function obtained using the LS estimator for the quasar pair-quasar at intermediate redshift is presented. The 2PCF for the entire sample, which has also been computed using a bootstrap method for resampling, is also included.  Bottom panel: The same two-point cross-correlation functions are displayed but in logarithmic scales.\label{fig:5}} 
\end{figure}
 
Now, we focus on the cross-correlation function on different scales, $2.2 \leq z\leq 2.8$ [cf.  Fig. (\ref{fig:5})]. 
The analysis reveals that the cross-correlation signal is considerably stronger for separations $s<20 \text{Mpc}h^{-1}$. In other words, quasar pairs within this range exhibit a more significant clustering tendency, suggesting a closer spatial relationship among these objects..  The cross-correlation function of (quasar pairs)$\times$(quasars) is found to be 5.98 times higher than that of the autocorrelation function of quasars when considering separations $s<20 \text{Mpc}h^{-1}$.  At large separations, both signals exhibit a similar decline, demonstrating a consistent behavior in the correlation functions. To gain insight into how the correlation length for the cross-correlation of quasar pairs with quasars compares to the quasar-quasar case, we can observe that the condition $\xi_{pq}(s_{0,pq})=1$ corresponds to a scale of approximately $s_{0,pq}\simeq 18.94$. In contrast, for the autocorrelation case, where $\xi_{qq}(s_{0,qq})=1$, the scale is around $s_{0,qq}\simeq 9.64$. The visual examination indicates a significant difference in the correlation lengths between these two scenarios, highlighting the influence of the cross-correlation on the spatial distribution of quasar pairs. \footnote{To grasp the significance of our earlier results, we can consider the formulation of the signal-to-noise ratio in the context of cross-correlation: $S/N \propto \sqrt{N_{q}N_{p}} \xi_{pq}(s)$. The latter relationship emphasizes that a nonzero pair detection is fundamentally encoded in the amplitude of the cross-correlation function, $\xi_{pq}(s)$, when juxtaposed against the background of the quasars \cite{menard}. Here, the interaction between the number of quasars $N_{q}$ and the number of potential counterparts $N_{p}$ amplifies the sensitivity of our measurements, allowing us to discern genuine correlations amidst the noise inherent in the quasar distribution.} As expected, the correlation length of the quasar pairs is greater than that of individual quasars.\\

We explored the implications of excluding quasar pairs from our sample and focused solely on the two-point correlation function derived from the modified dataset (autocorrelation). It allows us to better understand the characteristics and clustering behavior of the remaining quasars. It appears that the signal maintains a low level, similar to that observed in the full sample. Specifically, within the linear regime, we find that the parameters of the power law are $s_{0}=(10.84\pm 1.99)\text{Mpc}h^{-1}$ and the exponent is $\gamma=(1.36\pm 0.06)$. These results are consistent with our previous findings, reinforcing the robustness of our analysis.\\

\begin{figure}[htbp]
\includegraphics[width=\linewidth]{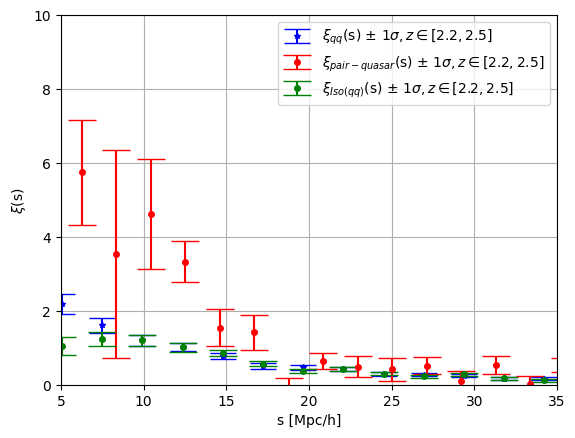}
\includegraphics[width=\linewidth]{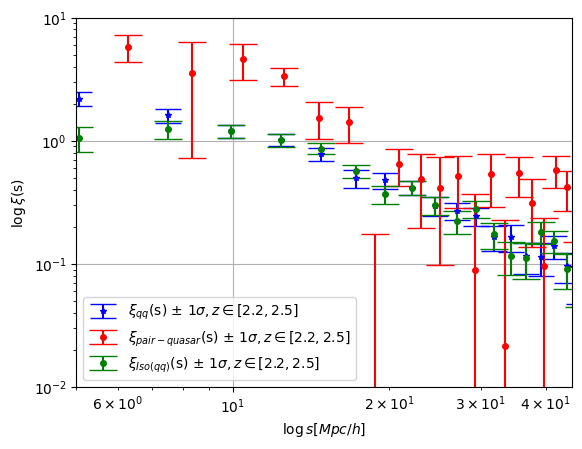}
\caption{Top panel: The two-point function is derived using the LS estimator for the cross-correlation between quasar pairs and individual quasars within the redshift interval $2.2 \leq z\leq 2.5$.  The 2PCF for the entire subsample is included, which has also been computed using the resampling bootstrap method, as well as for isolated quasars that do not belong to any pair. Bottom panel: the same two-point functions are presented in a logarithmic scale.\label{fig:6}} 
\end{figure}

Another important aspect to consider is the impact of splitting the subsample in the redshift range $2.2 \leq z\leq 2.8$ into two smaller subsets: $2.2 \leq z\leq 2.5$ and $2.5 \leq z\leq 2.8$. Such a partition allows us to investigate potential differences in clustering behavior and correlation properties between these two redshift intervals, providing a clearer understanding of how the quasar pair distribution evolves with redshift.  For the sake of brevity, we will present the correlation functions for the redshift range $2.2 \leq z\leq 2.5$, as these results are sufficient to support our conclusions. In examining the cross-correlation, we observe a notable variability in the residuals that correlates with the level of the independent variable.  We conclude that the assumption of constant variance does not hold with the help of the Breusch-Pagan algorithm. To address the latter issue, we proceed by performing a weighted linear regression analysis.  A weight coefficient of $R^{2}_{WLR}=0.80$ is obtained, along with the following parameter estimates: $\gamma=2.77\pm 0.44$ and $s_{0}=(16.27\pm 3.27)\text{Mpc}h^{-1}$. \footnote{For the other redshift partition ($2.5 \leq z\leq 2.8$), we encounter no significant variance issues, allowing us to proceed with a standard linear regression analysis. Indeed, when we conduct a comprehensive nonlinear fit for the initial eight bins, we obtain a correlation length given by $s_{0}=(16.00\pm 1.94)\text{Mpc}h^{-1}$ and the exponent is $\gamma=(1.65\pm 0.16)$.} All in all, we find that the cross-correlation function of quasar pairs with quasars exhibits a significantly stronger signal compared to both the isolated case (quasars that do not belong to pairs) and the standard quasar-quasar correlation. The difference is visually apparent in Fig. (\ref{fig:6}), where the enhanced clustering signal of the cross-correlation is clearly discernible. \\

\begin{figure}[htbp]
\includegraphics[width=\linewidth]{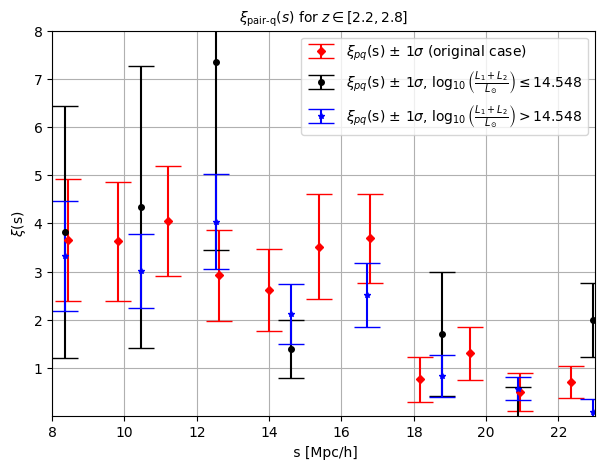}
\includegraphics[width=\linewidth]{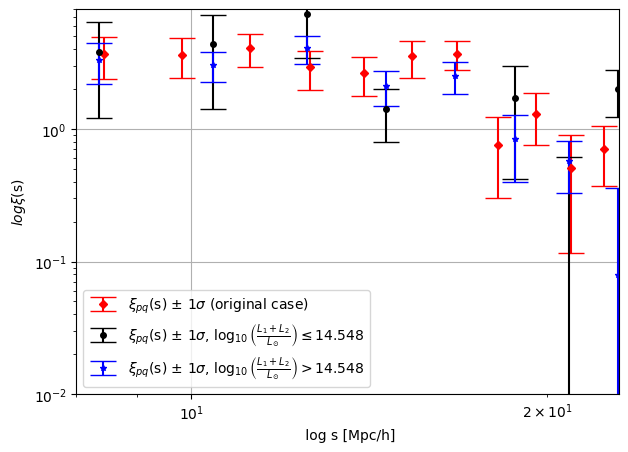}
\caption{Top panel: Comparison of two-point cross-correlation function obtained using the LS estimator for the quasar pair-quasar in the $2.2 \leq z\leq 2.8$ redshift window  is shown but using different partition criteria. Bottom panel: The same two-point cross-correlation functions are displayed using a logarithmic scale to visualize their slope.\label{fig:10}} 
\end{figure}

We suggest an alternative approach to validate that the observed cross-correlation is physically significant and not merely a consequence of the chosen redshift bins. It involves partitioning the original subsample within the redshift window $z \in [2.2, 2.8]$ into two distinct datasets, which are created by applying a cutoff on the logarithm of the sum of the luminosities. By adopting the aforesaid approach, we can further assess whether our results maintain physical relevance across different partitions bu using the luminosities of the quasar pairs as an input. In fact, Fig. (\ref{fig:10}) shows that the cross-correlation signal persists regardless of the cut-off or partitioning method employed. 
Although the error bars may appear larger, even with the same bootstrap method applied, the larger size is primarily associated with the reduced sample size after partitioning based on the following specified criterion,  $\log_{10}\big(L_{1} + L_{2}\big)/L_{\odot} \lessgtr 14. 548$.  Such observation underscores the resilience of the correlation between different approaches, reinforcing the validity of our findings despite the variations in data partitioning. As anticipated, linear regression may be less reliable in certain scenarios. For the case where $\log_{10}\big(L_{1} + L_{2}\big)/L_{\odot} > 14. 548$, the slope is found to be $\gamma= (2.00\pm 0.40)$ and the correlation length is $s_{0}= (16.00\pm 1.70)$, with a coefficient of determination $R^{2}\simeq 0.80$. In contrast, for the opposite case where $\log_{10}\big(L_{1} + L_{2}\big)/L_{\odot} \leq 14. 548$, the estimations yield $\gamma= (2.00\pm 0.63)$, $s_{0}= (16.00\pm 4.15)$, and $R^{2}\simeq 0.68$.   

These findings warrant further exploration, which will be undertaken in the next section through an investigation of the convergence CMB lensing signal \ref{lensing}. Additionally, we will assess whether the amplitude of the convergence signal is primarily influenced by the more luminous quasar pairs or the less luminous ones.


\section{CMB Lensing Signal} \label{lensing}
In this section, we study the association of mass with quasars and quasar pairs using CMB $\kappa$ convergence maps. Given the previous results obtained and the fact that the hosts of quasars are galaxies, we expect a higher $\kappa$ convergence signal in the neighborhood of quasar pairs compared to the vicinity of isolated quasars. Previous studies for similar samples have shown a lensing signal at $\kappa \sim 2 \cdot 10^3$ at angular distances below one degree \cite{Geach, Petter, Eltvedt}. Taking this into account, we follow the methodology of \cite{Toscano} and calculate radial convergence profiles using the CMB data products released in 2018 by the Planck Collaboration \cite{Planck_lensing}. We reconstruct the map from the spherical harmonic coefficients $k_{LM}$\footnote{\url{https://pla.esac.esa.int/\#home}}, using all the $L's$ available ($L_{Max}=4096$) and a smoothing scale of $0.2^{\circ}$, which corresponds roughly to 3.5 Mpc/h. We use the individual redshift of each quasar to project the angular scale of each object to its corresponding distance scale (in Mpc) in order to compare the lensing results with the correlation results obtained above. As in the previous section, we consider isolated quasars and those in pairs, applying the same criteria for the distinction of these two sub-samples.\\

\begin{figure} [htpb]
    \centering
    \includegraphics[width=\linewidth]{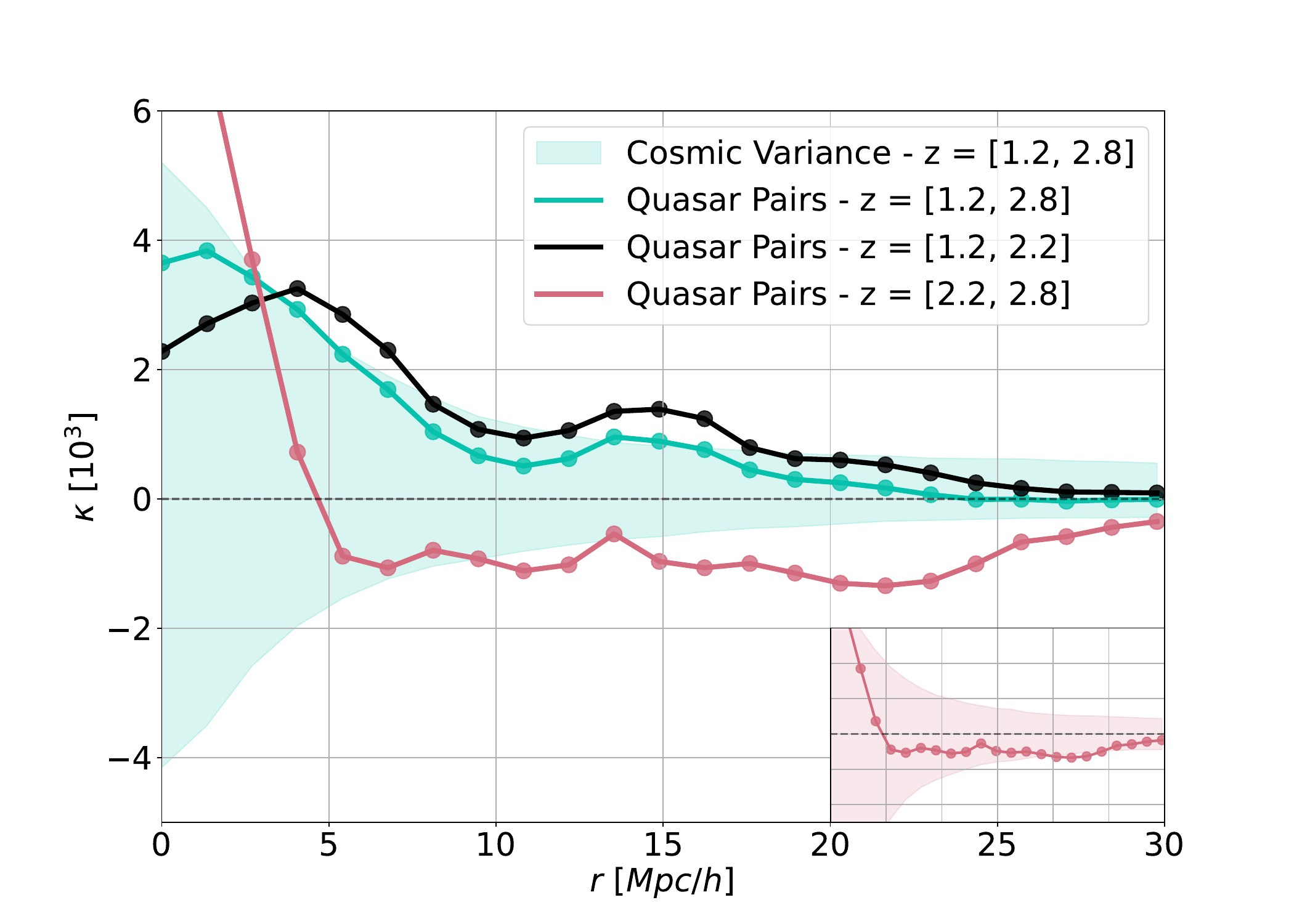}
    \caption{ Radial Convergence Profiles for the quasars sample in the redshift range $1.2 \leq z \leq 2.8$ divided into pairs of quasars in different redshift intervals. Inset Panel: Quasar pairs in the redshift range $z \in [2.2 , 2.8]$ with their corresponding cosmic variance, showing a negligible signal-to-noise ratio.}
    \label{fig:kappa_global}
\end{figure}

We conduct a general analysis of the CMB convergence signal across the  redshift interval $z \in [1.2, 2.8]$ (associated with the full sample) and estimate the cosmic variance within this range [see Fig. (\ref{fig:kappa_global})]. The signal appears to fall within the bounds of the cosmic variance estimation, however the quasar pairs in the redshift range of $[1.2,2.2]$ exhibit a larger amplitude compared to the overall sample and the cosmic variance; confirming the previous analysis on cross-correlation function quasar-quasar pairs [cf. Fig. (\ref{fig:11})]. This behavior is congruent with the maximum amplitude expected from the theory, taking into account that the source is located at $z \sim 1100$. Therefore, we further study the redshift range $z \in [1.2, 2.2]$ where $ 2126$ pairs were established in order to maximize the signal-to-noise ratio.\\

\begin{figure} [htpb]
    \centering
    \includegraphics[width=\linewidth]{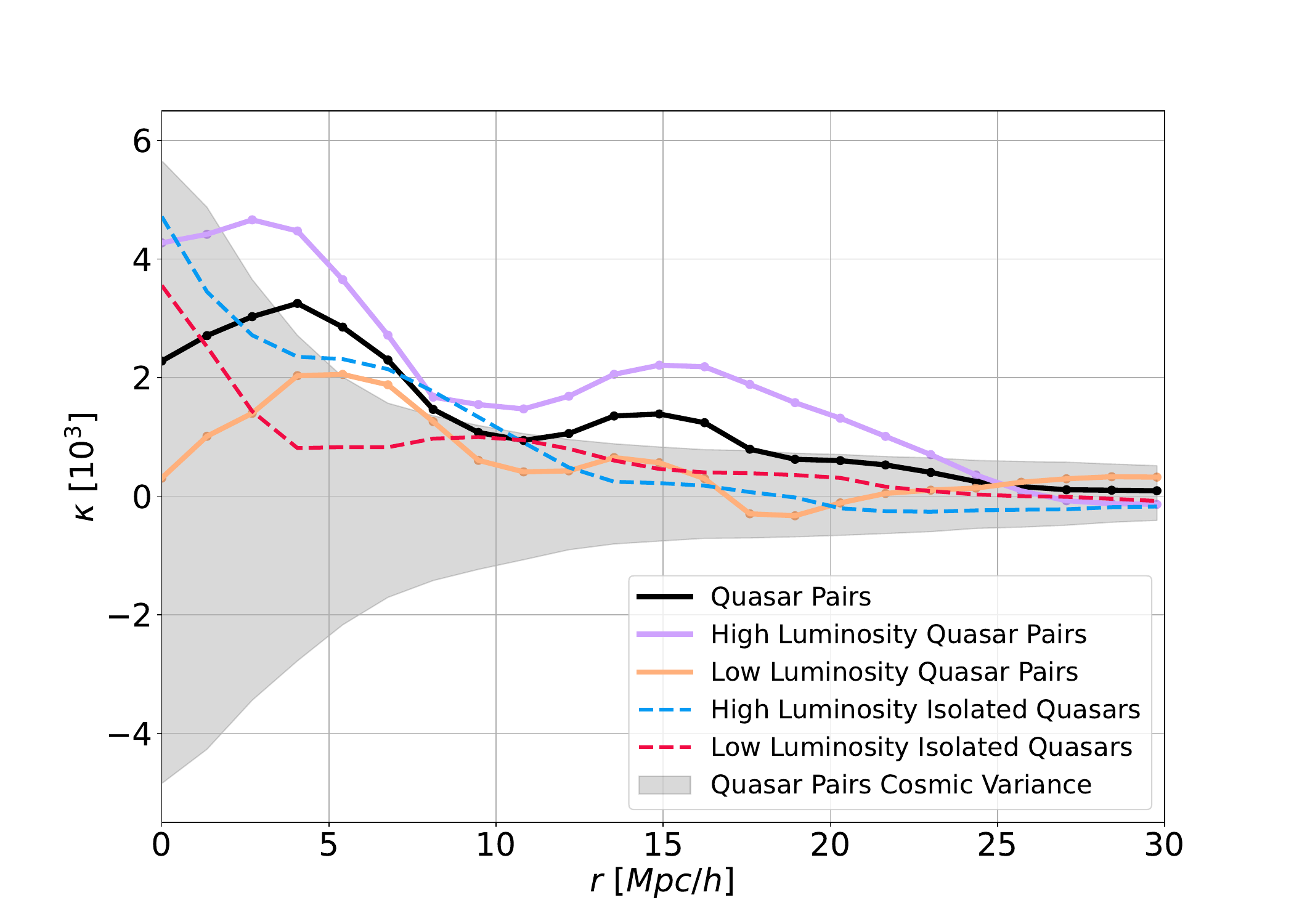}
    \caption{The radial convergence profiles for quasar pairs, categorized by luminosity levels (dotted lines for high and low luminosity), are shown in the redshift range $1.2 \leq z \leq 2.2$. 
     For comparison, the profiles of isolated quasars are shown with dashed lines. The results indicate that the amplitude of the radial convergence for quasar pairs consistently exceeds that of isolated quasars. Moreover, the high luminosity quasar pairs appear to contribute significantly to a signal that surpasses the expected cosmic variance}.
    \label{fig:kappa_qsos}
\end{figure}

Figure \ref{fig:kappa_qsos} shows the radial convergence profiles for isolated quasars and quasar pairs in this redshift range. For the isolated ones, we further divided the sample into two new subsets: high-luminosity and low-luminosity quasars, according to the median luminosity of the total sample. It is clear that even with this sub-sampling, there is no difference in signal for the isolated quasars, however, as in the case of the cross-correlation function, quasar pairs have a significantly higher signal than the isolated quasar sample. Furthermore, if we divide the pairs into high-luminescence and low-luminescence pairs taking into account the median of $L_1+L_2$, we find that the most luminous quasars provide the largest signal contribution to the total sample signal. This significance can be noticed if we consider random positions in the CMB map and estimate the cosmic variance of the sample of pairs, showing a high noise-signal relation. 


\section{Summary and Discussions}

In this study, we construct a sub-sample from the SDSS DR16 quasar catalog, focusing on a redshift range ($z \in [1.2, 2.8]$) where the sample exhibits a good completeness factor. We conducted a comprehensive analysis of this subsample, with particular emphasis on the detection of quasar pairs. To achieve this, we first calculated the projected distance in Mpc between the two members of each quasar pair.\\

We performed an extensive exploratory analysis of the luminosity distribution of the quasar pairs, examining how they distinguish themselves from both control and simulated pairs.  Moreover, we computed the cross-correlation function between the quasar pairs and the overall quasar population, revealing a distinctly amplified signal above the background of isolated quasars. We investigated how this correlation is influenced by various scenarios, including partitioning the redshift interval into two segments, separating the data according to the radio luminosity ratio $L_1/L_2$, and applying a cutoff in $\log_{10}(L_1+L_2)/L_{\odot}$. The main amplitude of the cross-correlation signal remained consistent under these conditions. We also addressed the observational capabilities for detecting quasar pairs using the SDSS camera, discussed the possibility of contamination of the quasar pair sample by foreground galaxies, and contrasted the numerical comoving density of the quasar pairs with that of the quasar subsample.\\ 

We have demonstrated that the CMB convergence signal remains within the bounds defined by cosmic variance estimates for the redshift interval $z\in [1.2, 2.8]$. When we refine our focus to a narrower redshift range, $z\in [1.2, 2.2]$, the CMB convergence signal associated with quasar pairs exhibits a heightened amplitude, surpassing not only the total quasar pair sample but also the estimated cosmic variance, as illustrated in Fig. (\ref{fig:kappa_global}). Furthermore, our analysis confirms that the amplitude of the cross-correlation between quasar pairs and individual quasars consistently exceeds that of the quasar auto-correlation within the same redshift range (see Fig. \ref{fig:11}). By integrating the cross-correlation function with the CMB convergence signal, we achieved a substantially enhanced detection rate of quasar pairs at lower redshifts, culminating in a total of $N_{p}=2126$.\\

We further validated our detection through an additional observational test, focusing on the amplitude of the CMB lensing signal associated with quasar pairs distinguished by high and low luminosities, using the mean luminosity as the cutoff between these two groups. Our analysis revealed that the most luminous pairs of quasars are the dominant contributors to the overall signal for the sample, as illustrated in Fig. \ref{fig:kappa_qsos}.\\

The present work could provide a valuable foundation for further investigating the characterization of the Large--scale structure by integrating the cross-correlation function obtained from quasar catalogs with an in-depth analysis of the CMB convergence lensing signal.

\begin{acknowledgements}
      We acknowledge the use of the \textit{SDSS DR16 catalog}   \url{https://www.sdss4.org/dr16/algorithms/qso_catalog}, \cite{Lyke}. We used several packages such as \textit{Astropy} \cite{astropy}, \textit{Corrfunc} \cite{corfun}, \textit{Numpy} \cite{np}, \textit{Matplotlib} \cite{plt}, \textit{Scipy} \cite{scipy},\textit{ Healpy} \cite{zonca}, \textit{HEALPix} \cite{healpix}, \textit{Pandas} \cite{pd}. The authors thank CNPq, FAPES, and Funda\c{c}\~ao Arauc\'aria for financial support. The authors would like to express their gratitude for the valuable comments provided by the referee, which significantly contributed to enhancing the quality and clarity of the paper.
\end{acknowledgements}

\section*{Data Availability}
The data that support the findings of this study will be available on reasonable request.

\end{document}